\documentclass[12pt]{article}

\RequirePackage{amsthm,amsmath,amsfonts,amssymb}
\RequirePackage[authoryear]{natbib}
\RequirePackage[colorlinks,citecolor=blue,urlcolor=blue]{hyperref}
\RequirePackage{graphicx}
\usepackage{geometry}
\usepackage{float}
\usepackage{array}
\newcolumntype{L}{>{\let\newline\\\arraybackslash\hspace{0pt}}m{4cm}}
\newcolumntype{P}{>{\centering\let\newline\\\arraybackslash\hspace{0pt}}m{2cm}}
\usepackage{multirow} 
\usepackage{algorithm}
\usepackage[noend]{algpseudocode}
\usepackage{enumitem}
\newtheorem{theorem}{Theorem}

\title{An Efficient Approach to Design Bayesian Platform Trials}

\author
{\small \textbf{Luke Hagar}$^{1,a,*}$, \textbf{Lara Maleyeff}$^{1,b,*}$, \textbf{Shirin Golchi}$^{1,c}$, \textbf{and Dick Menzies}$^{2,d}$\\
\footnotesize $^{1}$Department of Epidemiology, Biostatistics, and Occupational Health, \\ \footnotesize McGill University\\
\footnotesize $^{2}$McGill International Tuberculosis Centre, Research Institute of the \\ \footnotesize  McGill University Health Centre and Department of Medicine, McGill University\\
\bigskip
\footnotesize \textsuperscript{*}These authors contributed equally.\\
\footnotesize \textsuperscript{a}\textit{luke.hagar@mail.mcgill.ca} \quad
\textsuperscript{b}\textit{lara.maleyeff@mcgill.ca} \\
\footnotesize  \textsuperscript{c}\textit{shirin.golchi@mcgill.ca} \quad
\textsuperscript{d}\textit{dick.menzies@mcgill.ca}
}
\date{}
\begin{document}

\maketitle

\begin{abstract}
Platform trials evaluate multiple experimental treatments against a common control group (and/or against each other), which often reduces the trial duration and sample size. Bayesian platform designs offer several practical advantages, including the flexible addition or removal of experimental arms using posterior probabilities and the incorporation of prior/external information. Regulatory agencies require that the operating characteristics of Bayesian designs are assessed by estimating the sampling distribution of posterior probabilities via Monte Carlo simulation. It is computationally intensive to repeat this simulation process for all design configurations considered, particularly for platform trials with complex interim decision procedures. In this paper, we propose an efficient method to assess operating characteristics and determine sample sizes as well as other design parameters for Bayesian platform trials. We prove theoretical results that allow us to model the joint sampling distribution of posterior probabilities across multiple endpoints and trial stages using simulations conducted at only two sample sizes. This work is motivated by design complexities in the SSTARLET trial, an ongoing Bayesian adaptive platform trial for tuberculosis preventive therapies (ClinicalTrials.gov ID: NCT06498414). Our proposed design method is not only computationally efficient but also capable of accommodating intricate, real-world trial constraints like those encountered in SSTARLET. 
\end{abstract}

\section{Background}

Multi-arm multi-stage (MAMS) and platform trial designs allow multiple experimental treatments to be evaluated against a shared control and/or against each other within a single trial infrastructure \citep{magirr2012generalized,royston2003novel, bauer1999combining, stallard2003sequential}. These designs can offer efficiency gains compared to conducting a series of independent two-arm trials, both in terms of required sample size and trial duration. They also provide practical advantages, such as reduced need for repeated ethics and regulatory reviews, minimized downtime between trials, and greater continuity in staffing--helping to reduce turnover and training requirements.\citep{parmar2008speeding}. Classical MAMS designs assume that all arms are present from the start and that the allocation ratio remains fixed across arms and stages. However, in the context of platform trials—ongoing trial structures that evaluate evolving treatment landscapes—there is growing interest in the ability to add new experimental arms mid-trial. This flexibility allows for efficient comparison of newly emerging interventions using the same infrastructure and potentially shared control data \citep{burnett2024adding}. While appealing, such adaptations raise important statistical challenges, particularly in preserving strong control of the family-wise error rate (FWER) across multiple comparisons and time points. 

Recent work in the frequentist framework has extended the Dunnett framework and closed testing procedures to accommodate preplanned arm additions while maintaining rigorous error control \citep{greenstreet2024mams, dunnett1955multiple}. Bayesian group sequential designs offer a flexible alternative in which stopping rules are based on posterior probabilities that a treatment is effective \citep{spiegelhalter2004bayesian, berry2010bayesian}. For example, a Bayesian trial might stop early for efficacy if the posterior probability that the treatment effect exceeds a clinically relevant threshold is greater than 0.99. These decision rules can be calibrated via simulation to ensure that the design achieves prespecified operating characteristics, including the frequentist type I error rate and power. In settings with multiple treatment arms or multiple outcomes, such as co-primary efficacy and safety endpoints, Bayesian designs allow separate monitoring boundaries for each criterion, using models that capture outcome interdependencies and trade-offs \citep{thall1995bayesian}.

A key advantage of Bayesian methods is the ability to incorporate historical data via informative priors. One principled framework for this is the meta-analytic predictive (MAP) prior, which uses a hierarchical model to summarize data from past studies and generate a predictive prior for the control arm in the current trial. This approach allows uncertainty due to between-study heterogeneity to be explicitly accounted for, improving efficiency without overstating the historical evidence. When historical data are compatible with current observations, MAP priors can reduce required sample sizes and improve precision. However, if there is conflict between historical and current data, posterior inference may become biased. To address this, robust MAP priors include a weakly informative component that down-weights the historical contribution in the presence of conflict, preserving type I error while retaining the benefits of borrowing when appropriate \citep{neuenschwander2010summarizing, schmidli2014robust}.

Despite their flexibility and statistical rigor, Bayesian adaptive designs have seen limited adoption in confirmatory trials. The U.S. Food and Drug Administration (FDA) has issued guidance supporting Bayesian approaches in both medical device trials \citep{fdaguidance2010bayesian} and drug and biologic trials \citep{fdaguidance2019adaptive}, emphasizing the acceptability of Bayesian methods when supported by adequate pre-trial simulations. Still, many investigators remain hesitant. Barriers include a lack of training in Bayesian inference, limited availability of validated software, and unfamiliarity with simulation-based design evaluation. Practical challenges also arise when specifying prior distributions, especially in regulatory settings where skepticism about subjectivity remains. Furthermore, few published phase III trials serve as templates for Bayesian implementation, particularly in multi-arm or multi-outcome settings. This gap extends to phase II, though adaptive designs increasingly blur the line between phases by enabling seamless transitions within a single protocol.

Sample size determination (SSD) for Bayesian adaptive trials typically requires extensive simulations to estimate sampling distributions of posterior summaries across a variety of sample sizes and decision criteria \citep{wang2002simulation}. Design operating characteristics are assessed using these sampling distribution estimates. To reduce the computational burden of designing non-sequential trials, sampling distributions of posterior summaries have been efficiently modeled throughout the parameter space \citep{golchi2022estimating, golchi2024estimating} and the sample size space \citep{hagar2025economical}. \citet{hagar2025ssd} extended the latter framework that relies on estimating the sampling distribution of posterior summaries at only two sample sizes to sequential designs without response-adaptive randomization. However, the dynamic treatment allocation inherent to platform trials prompts response-adaptive randomization when decisions to add or remove experimental arms are based on the accrued data. Most recently, the framework to explore the sample size space was adapted to non-platform Bayesian trials with complex decision rules based on clustered data and multiple endpoints \citep{hagar2025clustered}. While this framework can be used to efficiently design various Bayesian trials, its functionality with platform trials is limited.

In this paper, we address SSD in a design with variable allocation ratios and a fixed delay between reaching interim recruitment thresholds and performing the corresponding analyses. This delay, common in global and logistically complex trials, is not accommodated by existing efficient SSD frameworks. We apply our proposed SSD method in the context of the SSTARLET trial, a cluster-randomized, multi-arm Bayesian adaptive trial evaluating short-course tuberculosis (TB) preventive therapy (ClinicalTrials.gov ID: NCT06498414). The trial includes a fixed delay between mid-trial arm addition and early stopping for futility, response-adaptive randomization induced by potential early stopping, and robust MAP priors for historical borrowing. We demonstrate how our proposed method maintains computational efficiency while preserving rigorous control of type I error  under real-world constraints. Our results support the broader adoption of simulation-calibrated Bayesian designs in global health trials where logistical realities demand both flexibility and efficiency.

The structure of this article is as follows. Section \ref{sec:SSTARLET} describes the SSTARLET trial, which serves as our motivating example for this work. In Section \ref{sec:methods}, we outline the complex design and decision criteria associated with our motivating example. Section \ref{sec:theory} develops theoretical results that substantiate our proposed SSD method for Bayesian adaptive platform trials, which efficiently estimates the sampling distribution of posterior probabilities at only two sample sizes. In Section \ref{sec:sim}, numerical studies confirm the strong performance of our method. We conclude in Section \ref{sec:disc} with a discussion of future work.

\section{Shorter and Safer Treatment Regimens for Latent Tuberculosis
(SSTARLET) Trial}\label{sec:SSTARLET}

TB remains a major global health challenge and has recently surpassed COVID-19 as the leading infectious cause of mortality worldwide \citep{who2022}. Although effective tuberculosis preventive treatments (TPT) are available, their impact on global TB incidence has been limited. To achieve meaningful reductions, TPT use would need to be massively expanded. However, current regimens are often too long and poorly tolerated to support widespread implementation.\citep{uplekar2015,dye2013}. To meet the United Nations General Assembly target of delivering TPT to 45 million people by 2028, there is an urgent need for shorter, safer, and more scalable regimens \citep{who2022}. The SSTARLET trial (Shorter and Safer Treatment Regimens for Latent Tuberculosis) was launched in response to this need. As a multi-arm, cluster-randomized Bayesian adaptive trial conducted in Canada and four additional countries with moderate to high TB incidence, SSTARLET is designed to evaluate shorter and safer TPT regimens using a platform design that enables interim decision making, mid-trial arm addition, and historical borrowing--all while navigating the logistical realities of large-scale implementation.

Participants will be randomized to one of several self-administered daily regimens: 4R10 (standard 4-month regimen of daily rifampin), 2R20 (double-dose rifampin for 2 months), 1LP (levofloxacin and rifapentine for 1 month), or an additional experimental arm to be added mid-trial. Randomization will initially occur in a 1:2:2 ratio across the 4R10, 2R20, and 1LP arms, respectively, to account for the greater prior information available for the 4R10 regimen. The additional experimental regimen will be added mid-trial, with 50\% of new participants allocated to the new arm and the remaining 50\% distributed equally among the remaining arms.

The trial’s primary outcome is the incidence of severe treatment-related adverse events (AEs) leading to permanent discontinuation, adjudicated by a blinded panel. Secondary outcomes include treatment completion (defined as taking $\geq 80\%$ of doses within 150\% of the intended time) and tolerability (self-reported symptoms but not severe enough to stop treatment).

An adaptive Bayesian framework will guide trial decisions, with non-inferiority margins of 4\% for AEs and 10\% for completion and tolerability. An interim analysis is planned once a prespecified number of patient outcomes become available (e.g., 700), with results expected after an additional 300 participants are enrolled due to outcome lag (e.g., at 1000 enrolled). Regimens not meeting predefined safety or efficacy thresholds may be dropped at interim, and their planned sample sizes reallocated proportionally across 4R10 and the remaining experimental arms to preserve overall power. The third experimental regimen is introduced when the interim analysis is triggered (e.g., at 700 enrolled), but decisions about dropping one or more arms are not made until interim results become available (e.g., at 1000 enrolled). The final and interim sample sizes are subject to tuning, with 26 months of planned follow-up for incident tuberculosis disease. This adaptive design improves efficiency by focusing enrollment on regimens most likely to meet safety and efficacy criteria, while maintaining rigorous oversight.

Although SSTARLET, in principle, follows a platform design where the trial may continue perpetually beyond the second analysis, we focus on the specific window that includes only two analyses. Therefore, what we refer to as the final analysis is in fact the final analysis of this specific window in the trial.

\begin{table}[ht]
\caption{Summary of AE rates, Beta prior parameters, from previous studies}
\centering
\label{tab:map_prior}
\renewcommand{\arraystretch}{1.2}
\begin{tabular}{|l|c|c|c|}
\hline
\textbf{Study} & \textbf{Regimen} & \textbf{Events/Total (\%)} & $(\alpha, \beta)$ \\
\hline
\cite{ruslami2024high} & 2R20 & 8/441 (1.8\%) & (9, 434) \\
\hline
\cite{ruslami2024high} & 4R10 & 15/440 (3.4\%) & (16, 426) \\
\hline
\cite{menzies2018} & 4R10 & 15/422 (3.6\%) & (16, 408) \\
\hline
\cite{menzies2008adverse} & 4R10 & 15/393 (3.8\%) & (16, 379) \\
\hline
\cite{menzies2004treatment} & 4R10 & 2/58 (3.4\%) & (3, 57) \\
\hline
\end{tabular}
\end{table}

To inform the prior distribution for the AE rate in the 4R10 and 2R20 arms of the SSTARLET trial, we construct informative and robust MAP priors using data from four previous clinical trials. Each of the prior studies used to inform these priors included adult or adolescent participants who were given identical regimens to the SSTARLET trial, and all used consistent AE definitions adjudicated by independent review panels. These features, alongside similar eligibility criteria, trial settings, and study populations, support the assumption of exchangeability with SSTARLET and justify the use of a MAP prior based on their results. Beta-Binomial priors were constructed from each of these sources using a weakly informative approach, and the resulting parameters are summarized in Table~\ref{tab:map_prior}.

\section{Methods}\label{sec:methods}

\subsection{Notation}
\label{sec:notation}
Let $k \in \{1,2,3\}$ index the trial outcomes, with $k = 1$ representing AEs, $k = 2$ non-completion, and $k = 3$ non-tolerability. Let $j \in \{0,1,2,3\}$ index treatment arms, where $j=0$ denotes the control (4R10) regimen and $j=1,2,3$ are experimental arms, with $j=1$ representing 2R10, $j=2$ representing 1LP, and $j=3$ representing the third added arm. For each treatment arm $j$ and outcome $k$, we let $\theta_{jk}$ denote the true event probability.

Let $n_1 = n$ and $n_2$ be the sample sizes at the interim and final analyses, respectively, with $n_2 = c_2n$ for a constant $c_2$ (see Section~\ref{sec:methods.3} for more detail). We let $\mathcal{D}_{n_tjk}$ denote the data available at analysis $t = 1, 2$ for comparing outcome $k$ between experimental arm $j \in \{1,2,3\}$ and the control arm $j = 0$. Each $\mathcal{D}_{n_tjk}$ includes only the outcome-$k$ data for arms $j$ and 0 available at sample size $n_t$; that is, comparisons are outcome-specific and pairwise. We define $\mathcal{D}_n = \{\mathcal{D}_{n_tjk} : t = 1, 2;\, j = 1, 2, 3;\, k = 1, 2, 3\}$ as the full collection of data used for interim and final decisions.

\subsection{Trial Hypotheses and Decision Criteria}\label{sec:methods.1}

For each outcome $k$ and experimental arm $j \in \{1,2,3\}$, we define non-inferiority hypotheses comparing arm $j$ to the control arm as:
\[
H_{0,jk}: \theta_{jk} - \theta_{0k} \geq \delta_{m,k} 
\quad \text{vs.} \quad
H_{1,jk}: \theta_{jk} - \theta_{0k} < \delta_{m,k},
\]

\noindent where $\delta_{m,k}$ is the non-inferiority margin for outcome $k$. The subscript $m$ in $\delta_{m,k}$ emphasizes that this is a fixed margin constant, not estimated from data. For this trial, we set $\delta_{m,1} = 0.04$ for AEs, and $\delta_{m,2} = \delta_{m,3} = 0.10$ for non-completion and non-tolerability.

At the interim analysis, an experimental arm is dropped if $\tau_{1jk}(\mathcal{D}_{n_1jk}) = \Pr(\theta_{jk} - \theta_{0k} \ge \delta_{m,k} \mid \mathcal{D}_{n_1jk}) > \gamma_k$ for any $k$, i.e., if sufficient evidence is available for the inferiority of arm $j$ to the control arm. Independently of the study results, a new experimental treatment $j=3$ will be added when interim enrollment is met. Final non-inferiority decisions are made using outcome $k = 1$ only, based on whether $\tau_{2j1}(\mathcal{D}_{n_2}) = \Pr(\theta_{j1} - \theta_{01} < \delta_{m,1} \mid \mathcal{D}_{n_2j1}) > \kappa_1$ for arms not previously dropped.

\subsection{Statistical Model and Posterior Inference}\label{sec:methods.2}

Let $y_{jk} \sim \text{Binomial}(n_{jk}, \theta_{jk})$ denote the number of participants in arm $j$ who experience outcome $k$, where $n_{jk}$ is the number of participants in that group. Each outcome is modeled independently using a binomial likelihood with a conjugate Beta prior. Although participants in this study are enrolled in household clusters, we analyze the data at the individual level.  A recent study with a similar population found that 87\% of clusters (1,001/1,154) were single-member households \citep{ruslami2024high}, suggesting a minimal clustering effect. For all $(j,k)$ pairs except $(j\in \{0,1\}, k=1)$, we use uniform priors:
\[
\theta_{jk} \sim \text{Beta}(1,1).
\]

To robustly incorporate historical information for the AE outcome in the control and 2R20 arms (\( j \in \{0,1\}, k = 1 \)), we specify a prior for the event probability \( \theta_{j1} \) as a mixture of an informative prior derived from historical randomized trials and a uniform component. Specifically, the prior is given by
\[
\theta_{j1} \sim w_1 \sum_{h=1}^{H} w_h \, \text{Beta}(a_h, b_h) + (1 - w_1)\, \text{Beta}(a_0, b_0),
\]
where \( H \) is the number of historical data sources, \( w_1 \in [0, 1] \) is the overall weight assigned to the informative portion of the prior, and \( w_h \) denotes the relative weight of each informative Beta component, normalized such that \( \sum_{h=1}^H w_h = 1 \). The parameters \( a_h \) and \( b_h \) are chosen to reflect the strength of evidence from each historical source, while the weakly informative component is specified as \( \text{Beta}(a_0, b_0) \), typically with \( a_0 = b_0 = 1 \) to yield a uniform prior.

To ensure robustness to prior-data conflict, we update the weights in the mixture using marginal likelihoods based on newly observed binomial data. Suppose we observe \( y \) AEs among \( n \) participants in arm $j$. Let $L_h(y \mid n)$ be the marginal likelihood of the data under the \( h \)th informative component and $L_0(y \mid n)$ be the marginal likelihood under the weakly informative component. The posterior weight assigned to the $h$th informative component is:
\[
w_h^{\text{post}} = \frac{w_1 w_h L_h(y \mid n)}{w_1 \sum_{h'=1}^H w_{h'} L_{h'}(y \mid n) + (1 - w_1) L_0(y \mid n)}.
\]

This formulation ensures that the contribution of the historical data to the posterior is down-weighted when there is prior-data conflict, as indicated by lower marginal likelihoods for the informative components. The updated posterior distribution for \( \theta_{j1} \) is then a mixture of updated Beta distributions for $j \in \{0,1\}$:
\[
\theta_{j1} \mid y, n \sim \sum_{h=1}^{H} w_{h}^{\text{post}} \, \text{Beta}(a_h + y, b_h + n - y) + \left(1 - \sum_{h=1}^{H} w_{h}^{\text{post}} \right) \text{Beta}(a_0 + y, b_0 + n - y).
\]
This robust updating scheme, often referred to as a mixture of experts or robust MAP prior, balances the use of historical evidence with a safeguard against miscalibration, allowing data from the current trial to override the informative prior when warranted.
We set \( w = 0.5 \) as is recommended in practice \citep{schmidli2014robust}.

Posterior inference for all other \( \theta_{jk} \) proceeds by computing its posterior distribution via conjugacy:
\[
\theta_{jk} \mid y_{jk} \sim \text{Beta}(\alpha_{jk}^{\text{post}}, \beta_{jk}^{\text{post}}),
\]
where \( \alpha_{jk}^{\text{post}} = \alpha_{jk}^{\text{prior}} + y_{jk} \) and \( \beta_{jk}^{\text{post}} = \beta_{jk}^{\text{prior}} + n_{jk} - y_{jk} \).

\subsection{Design Criteria}\label{sec:methods.3}

Our new SSD method proposed in Section \ref{sec:theory} accommodates varying randomization ratios, arm addition, and a fixed delay between accrual and interim decision making for experimental arms.  To motivate the method, we consider the setting of the SSTARLET trial; full design details are provided in Section~\ref{sec:SSTARLET}. 

Let \(n = n_1\) denote the number of participants whose outcomes will be used at the interim analysis. Due to a fixed lag between enrollment and outcome availability, the interim analysis is conducted only after \(n + 300\) participants have been enrolled, though it is based solely on outcomes from the first \(n\) participants. At the point when \(n\) participants have been accrued, a third experimental arm is added. From that point, 50\% of new participants are allocated to the new arm, and the remaining 50\% are divided equally among the control and the two original experimental arms. This allocation continues until the interim analysis results are available. 

To define final sample sizes for power calculations, we introduce a constant \( c_2 \), such that the total number of participants with observed outcomes at the final analysis is \(n_2 = c_2 n\) and the second stage (between interim enrollment and the final analysis) includes \((c_2 - 1)n\) participants. The first 300 of these are accrued before interim decisions are implemented. The remaining \((c_2 - 1)n - 300\) participants are enrolled post-interim; half are assigned to arm 3, and the other half are allocated equally among the arms \(j \in \{0,1,2\}\) that remain under evaluation.

As a result of this varying allocation ratio, the resulting sample sizes differ across four cases corresponding to the active arms—that is, the subsets of experimental arms that remain in the trial after the interim analysis. Let \( \mathcal{S} \in \left\{ \emptyset,\ \{1\},\ \{2\},\ \{1,2\} \right\} \) denote the set of active experimental arms after the interim analysis at sample size \(n\). Here, arm \(j=1\) corresponds to 2R20, arm \(j=2\) to 1LP, and arm \(j=3\) is the third experimental regimen introduced before the interim analysis regardless of outcomes. For example, \( \mathcal{S} = \{1,2\} \) implies that both 2R20 and 1LP are retained, while \( \mathcal{S} = \emptyset \) indicates that both were dropped and only the newly added arm proceeds to the final analysis.

For illustration, we consider the case where arm 2 is dropped. In this case, no further patients are randomized into arm 2 after the interim decision, and a greater proportion of patients are allocated to arm 1 and the control group. The control and arm 1 groups receive \( (0.2 + (c_2-1)/4) n - 25 \) and \( (0.4 + (c_2-1)/4) n - 25 \) participants, respectively; these sample sizes are non-proportional linear functions of $n$. Arm 3 accumulates no stage 1 observations, 150 patients from the fixed accrual window, and half of the second stage accrual, resulting in a sample size of \( 0.5 (c_2-1) n \). Appendix A of the online supplement details the final sample sizes under this variable allocation scheme for all elements in the set $\mathcal{S}$. 

We build upon the framework in \citet{hagar2025ssd} to assess the operating characteristics of the design. While \citet{hagar2025ssd} did not consider variable allocation ratios, their theoretical results can be directly applied to the case where the final sample size for each arm under all decision scenarios is \emph{proportional} to the sample size $n$. However, as demonstrated in Appendix A, in SSTARLET this proportionality is not attained when an experimental arm is dropped due to the fixed delay between the time that the interim analysis is triggered and the time of interim decisions. To accommodate such predictable delays in designs like that of SSTARLET, we require further theoretical developments. Before we develop this theory in Section \ref{sec:theory}, we overview how the sample size and decision criteria impact the operating characteristics of the SSTARLET trial.

To evaluate the operating characteristics under this design, we consider the joint sampling distribution of posterior probabilities, denoted \( \boldsymbol{\tau}(\mathcal{D}_n) \). Specifically,
we consider the sampling distribution of
\[
\boldsymbol{\tau}(\mathcal{D}_n) = 
\begin{bmatrix}
\boldsymbol{\tau}_1(\mathcal{D}_{n}) \\
\boldsymbol{\tau}_2(\mathcal{D}_{n})
\end{bmatrix},
\]

\noindent where  $\boldsymbol{\tau}_1(\mathcal{D}_{n_1}) = \left\{ \tau_{1jk}(\mathcal{D}_{njk}) : j \in \{1,2\},\ k \in \{1,2,3\} \right\}$ is the vector of interim posterior probabilities comparing 2R20 and 1LP to 4R10 across outcomes, and $\boldsymbol{\tau}_2(\mathcal{D}_n) = \left\{ \tau_{2jk}^{(s)}(\mathcal{D}_{n_2jk}) : j \in \{3\} \cup s,\ k \in \{1,2,3\}, s \in \mathcal{S}\right\}$ is the set of final posterior probabilities for each arm-outcome combination, conditional on active set $s$. In the actual trial, we only observe $\left\{ \tau_{2jk}^{(s)}(\mathcal{D}_n) : j \in \{3\} \cup s,\ k \in \{1,2,3\}\right\}$ for a single active set $s \in \mathcal{S}$; however, we must consider all components of $\boldsymbol{\tau}_2(\mathcal{D}_n)$ for design purposes. To prompt general notation, we let $L$ be the dimension of $\boldsymbol{\tau}(\mathcal{D}_n)$. 

We view the sampling distribution of $\boldsymbol{\tau}(\mathcal{D}_n)$ as arising from a probability model \( \Psi \) over the true model parameters \( \boldsymbol{\theta} \) for all arms, which differs from the analysis prior \( p(\boldsymbol{\theta}) \). The model \( \Psi \) thus serves as a design prior \citep{de2007using, brutti2014bayesian}. To align with how the SSTARLET trial is designed, we consider degenerate models $\Psi$ that do not incorporate uncertainty about the parameters \(\boldsymbol{\theta}\) in this paper. Guidance on accommodating non-degenerate models $\Psi$ is provided in \citet{hagar2025ssd}. 

For each replicate \( r = 1, \dots, R \), synthetic trial data \( \mathcal{D}_{n,r} \) are generated based on \( \boldsymbol{\theta} \sim \Psi \). The vector of posterior probabilities \( \boldsymbol{\tau}(\mathcal{D}_{n,r}) \) is then computed under the fixed analysis model for scenarios corresponding to all possible active sets. Across all \(R\) replicates, the resulting collection \( \{\boldsymbol{\tau}(\mathcal{D}_{n,r})\}_{r=1}^R \) forms an empirical approximation to the joint sampling distribution of \( \boldsymbol{\tau}(\mathcal{D}_n) \) under the model $\Psi$. This sampling distribution enables marginal estimation of power and the type I error rate for the adaptive design. In particular, power for experimental arm $j = 1, 2, 3$ can be expressed as the probability of correctly declaring non-inferiority for arm $j$, marginalizing over all possible active sets \( s \in \mathcal{S} \):

\begin{equation}\label{eq:power.true}
    \nu_j\left(\boldsymbol{\tau}\left(\mathcal{D}_{n}\right)\right) = \sum_{s \in \mathcal{S}} P(S = s) 
    P\left( \tau_{2j1}^{(s)}(\mathcal{D}_{n}) > \kappa_1 \mid S = s \right),
\end{equation}

\noindent where \( P(S=s) \) is the probability of observing each active set at the interim analysis and $\kappa_1$ refers to the inferiority cutoff common to all experimental arms. If arm $j \notin \{3\} \cup s$, then the conditional probability in (\ref{eq:power.true}) is 0. 

We empirically approximate power in (\ref{eq:power.true}) using the Monte Carlo replicates as

\begin{equation}\label{eq:power.est}
\widehat{\nu}_j\left(\left\{\boldsymbol{\tau}\left(\mathcal{D}_{n,r}\right)\right\}_{r=1}^R\right) \approx \frac{1}{R} \sum_{r=1}^R \mathbf{1}\left\{ \tau_{2j1}^{(s^{(r)})}(\mathcal{D}_{n,r}) > \kappa_1 \right\},
\end{equation}

\noindent where \( s^{(r)} \) is the active set determined at interim in replicate \(r\). The indicator function in (\ref{eq:power.est}) is defined to be 0 if $j \notin \{3\} \cup s^{(r)}$. We define theoretical power in (\ref{eq:power.true}) and empirical power in (\ref{eq:power.est}) under a model $\Psi_1$ such that $\{H_{1,j1}\}_{j=1}^3$ are true. Power generally increases with the sample size $n$, so we aim to choose the smallest sample size for the SSTARLET trial that attains pre-specified power criteria.

The FWER can be expressed as the probability of incorrectly declaring non-inferiority for \emph{at least one} retained treatment, marginalizing over all possible active sets \( s \in \mathcal{S} \):
\begin{equation}\label{eq:t1e.true}
    \nu_0\left(\boldsymbol{\tau}\left(\mathcal{D}_{n}\right)\right) = \sum_{s \in \mathcal{S}} P(S = s) \sum_{j \in \{3\} \cup s} 
    P\left( \tau_{2j1}^{(s)}(\mathcal{D}_{n}) > \kappa_1 \mid S = s \right).
\end{equation}
We empirically approximate the FWER in (\ref{eq:t1e.true}) using the Monte Carlo replicates as
\begin{equation}\label{eq:t1e.est}
\widehat{\nu}_0\left(\left\{\boldsymbol{\tau}\left(\mathcal{D}_{n,r}\right)\right\}_{r=1}^R\right) \approx \frac{1}{R} \sum_{r=1}^R \sum_{j \in \{3\} \cup s^{(r)}} \mathbf{1}\left\{ \tau_{2j1}^{(s^{(r)})}(\mathcal{D}_{n,r}) > \kappa_1 \right\},
\end{equation}
under a model $\Psi_0$ where $\{H_{0,j1}\}_{j=1}^3$ are instead true. The interim and final decision thresholds--$\{\gamma_k\}_{k=1}^3$ and $\kappa_1$--can be chosen to bound the FWER of the SSTARLET trial.

\section{A Method for Sample Size Determination}\label{sec:theory}

\subsection{Theory}

We motivate our SSD method proposed in Section \ref{sec:theory.2} by constructing a proxy to the joint sampling distribution of posterior probabilities. This theoretical proxy substantiates our methodology. However, our methods do not directly use these proxies and instead estimate the true sampling distribution of $\boldsymbol{\tau}(\mathcal{D}_n)$ by simulating synthetic trials and approximating posterior probabilities as described in Section \ref{sec:methods.3}. Our proxies are predicated on large-sample conduits for the data as well as asymptotic posterior approximations that are based on the Bernstein-von Mises (BvM) theorem \citep{vaart1998bvm}. 

The quantity for which we consider data conduits and the posterior distribution is $\boldsymbol{\delta} = \boldsymbol{\delta}(\boldsymbol{\theta})$, where the $i$th component of this vector $\delta_i$ is the estimand that corresponds to the $i$th component of $\boldsymbol{\tau}(\mathcal{D}_n)$ for $i = 1, \dots, L$. All estimands in the SSTARLET trial, for instance, are differences in Bernoulli proportions. Because $\boldsymbol{\tau}(\mathcal{D}_n)$ is indexed over all possible active sets, a given estimand can appear multiple times in $\boldsymbol{\delta}$. For example, the final posterior probability comparing AE rates in the control and arm 3 will appear in each of the four active sets, and thus will be represented four times in $\boldsymbol{\tau}(\mathcal{D}_n)$. The general notation presented in this subsection allows us to accommodate a broad range of platform trials with varied active sets defined across any number of interim analyses. 

The data conduit that we consider for $\boldsymbol{\delta}$ is the joint maximum likelihood estimate $\hat{\boldsymbol{\delta}}^{_{(n)}}$ that is independent of the interim decisions and indexed by the total sample size at the first analysis $n$. We suppose that the regularity conditions for the asymptotic normality of the maximum likelihood estimator (MLE) in Theorem 5.39 of \citet{vaart1998bvm} are satisfied. Under those conditions, the sampling distribution of the MLE $\hat{\boldsymbol{\delta}}^{_{(n)}}$ is
  \begin{equation}\label{eq:joint.mle}
  \hat{\boldsymbol{\delta}}^{_{(n)}} \sim \mathcal{N}\left(\boldsymbol{\delta}^{*} , \boldsymbol{\Sigma}(n)\right),
\end{equation}
where $\boldsymbol{\delta}^{*}$ is the value of $\boldsymbol{\delta}(\boldsymbol{\theta})$ under the model $\Psi$, $\boldsymbol{\Sigma}(n) = n^{-1}\boldsymbol{\Sigma} + \mathcal{O}(1/n^2)$, and $\boldsymbol{\Sigma}$ is a matrix related to the inverse Fisher information. In Appendix B of the online supplement, we show that $\boldsymbol{\Sigma}(n) = n^{-1}\boldsymbol{\Sigma} + \mathcal{O}(1/n^2)$ for trials like SSTARLET in which the sample sizes for all arms in all stages are linear functions of $n$. The matrix $\boldsymbol{\Sigma}$ accounts for the dependence between estimands over stages of the trial \citep{jennison2000group} and across endpoints. Since we only use the result in (\ref{eq:joint.mle}) for theoretical purposes, we need not obtain  $\boldsymbol{\Sigma}$ to use our method proposed in Section \ref{sec:theory.2}. 

The conditions for the BvM theorem in Theorem 10.1 of \citet{vaart1998bvm} require that the prior distribution for $\boldsymbol{\delta}$ is continuous with positive density at $\boldsymbol{\delta}^{*}$. When these conditions are satisfied as in the SSTARLET trial, the limiting posterior distribution of $\delta_i$ is
  \begin{equation}\label{eq:bvm}
  \mathcal{N}\left(\hat{\delta}^{_{(n)}}_i, \Sigma_{i, i}(n)\right),
\end{equation}
where $\Sigma_{i, i}(n)$ is the $(i,i)$-entry of $\boldsymbol{\Sigma}(n)$ from (\ref{eq:joint.mle}). Our proxy to the sampling distribution of $\boldsymbol{\tau}(\mathcal{D}_n)$ substitutes the MLEs generated via (\ref{eq:joint.mle}) into the posterior distribution in (\ref{eq:bvm}). This substitution yields the following approximation to the $i$th component of $\boldsymbol{\tau}(\mathcal{D}_n)$:
      \begin{equation}\label{eq:proxy}
\tau^{_{(n)}}_{i} = 
   \Phi\left(\dfrac{\delta_{m, i} - \hat{\delta}^{_{(n)}}_{i}}{\sqrt{\Sigma_{i, i}(n)}}\right), 
\end{equation} 
where $\delta_{m, i}$ is the non-inferiority margin corresponding to row $i$ of $\boldsymbol{\tau}(\mathcal{D}_n)$. While the total variation distance between such proxy sampling distributions and the true sampling distribution of $\boldsymbol{\tau}(\mathcal{D}_n)$ converges in probability to 0 as $n \rightarrow \infty$ \citep{hagar2024fast}, this proxy sampling distribution based on asymptotic theory only motivates our result in Theorem \ref{thm1}. 

Theorem \ref{thm1} explores how conditional quantiles of our proxy sampling distribution change as deterministic functions of $n$. We let $\tau^{_{(n)}}_{i, u}$ be the $u$-quantile of the sampling distribution of $\tau^{_{(n)}}_{i}$ in (\ref{eq:proxy}) induced by the sampling distribution of $\hat{\boldsymbol{\delta}}^{_{(n)}}$ in (\ref{eq:joint.mle}). Theorem \ref{thm1} guarantees that the logits of $\tau^{_{(n)}}_{i, u_i} ~|~\{\tau^{_{(n)}}_{i', u_{i'}}\}_{i' = 1}^{i-1}$ are approximately linear functions of $n$ for all $i \in \{1, \dots, L\}$. For $i = 1$, the conditioning set $\{\tau^{_{(n)}}_{i', u_{i'}}\}_{i' = 1}^{i-1}$ is empty. As demonstrated in Appendix B, the dependence structure between $\{\tau^{_{(n)}}_{i}\}_{i=1}^{L}$ is determined by the matrix $\boldsymbol{\Sigma}(n)$. Because these conditional quantiles change linearly with the sample size, we can model the entire joint proxy sampling distribution using linear functions of $n$. We later adapt this result to estimate the operating characteristics of platform trials across a broad range of sample sizes by estimating the true sampling distribution of $\boldsymbol{\tau}(\mathcal{D}_n)$ at only two values of $n$.

    \begin{theorem}\label{thm1}
     We suppose the conditions for the BvM theorem and those for the asymptotic normality of the MLE in \citet{vaart1998bvm} are satisfied. Define $\emph{logit}(x) = \emph{log}(x) - \emph{log}(1-x)$. Let the sample sizes for all treatments in all trial stages be linear functions of $n$ such that $\boldsymbol{\Sigma}(n)$ in (\ref{eq:joint.mle}) is $ n^{-1}\boldsymbol{\Sigma} + \mathcal{O}(1/n^2)$. Denote  the $(i,i)$-entry of the matrix $\boldsymbol{\Sigma}$ as $\Sigma_{i,i}$. Consider a given set of quantile levels $\{u_{i}\}_{i=1}^{L} \in [0,1]^{L}$. For $i \in \{1, \dots, L\}$, the conditional quantiles $\tau^{_{(n)}}_{i, u_i} ~|~\{\tau^{_{(n)}}_{i', u_{i'}}\}_{i' = 1}^{i-1}$ are such that
$$\lim\limits_{n \rightarrow \infty} \dfrac{d}{dn}~\emph{logit}\left(\tau^{_{(n)}}_{i, u_i} ~|~\{\tau^{_{(n)}}_{i', u_{i'}}\}_{i' = 1}^{i-1}\right)= (0.5 - \mathbb{I}\{\delta^*_i \ge \delta_{m,i}\})\times \dfrac{(\delta_{m,i} - \delta^*_i)^2}{\Sigma_{i,i}}.$$ 
\end{theorem}

We prove Theorem 1 in Appendix B. \citet{hagar2025ssd} put forward a similar result under the constraint that $\boldsymbol{\Sigma}(n)= n^{-1}\boldsymbol{\Sigma}$ when the sample sizes for all treatments in all stages are \emph{proportional} to $n$.  We emphasize that their result cannot be applied to design SSTARLET or similar platform trials. Theorem \ref{thm1} is framed in the context of one-sided hypotheses, but we could also apply our SSD framework with equivalence tests (see e.g., \citet{hagar2025ssd}).

We now consider the practical implications of Theorem \ref{thm1}. The limiting derivative is a constant that does not depend on $n$ nor the conditional quantile levels $\{u_{i}\}_{i=1}^{L}$. Thus, the limiting derivatives of the logits of $\tau^{_{(n)}}_{i, u_i}$ and $\tau^{_{(n)}}_{i, u_i}~|~\{\tau^{_{(n)}}_{i', u_{i'}}\}_{i' = 1}^{i-1}$ are the same. For $i \in \{1, \dots, L\}$, the linear approximation to the conditional quantile, $\text{logit}\left(\tau^{_{(n)}}_{i, u_i} ~|~\{\tau^{_{(n)}}_{i', u_{i'}}\}_{i' = 1}^{i-1}\right)$, as a function of $n$ is a good global approximation for sufficiently large sample sizes. This linear approximation is also locally suitable for a range of smaller sample sizes. In Section \ref{sec:theory.2}, we adapt this linear trend in the proxy sampling distribution to flexibly model the sampling distribution of  $\boldsymbol{\tau}(\mathcal{D}_n)$ using linear functions of $n$ when simulating synthetic trials. Although the proxy sampling distribution is predicated on asymptotic results, we illustrate the good performance of our SSD procedure with finite sample sizes $n$ in Section \ref{sec:sim}.

\subsection{Implementation for SSTARLET}\label{sec:theory.2}

   We generalize our result from Theorem \ref{thm1} to develop an SSD procedure for Bayesian platform trials. We detail this procedure in Algorithm \ref{alg.star} using the context of the SSTARLET trial and its decision criteria. Readers are directed to \citet{hagar2025ssd} for more general guidance on SSD in Bayesian adaptive trials. Algorithm \ref{alg.star} requires that we estimate the sampling distribution of posterior probabilities by simulating synthetic trial data at only two values of $n$: $n_{a}$ and $n_b$. The initial sample size for the interim analysis $n_a$ can be selected based on the anticipated budget for the trial. In Algorithm \ref{alg.star}, we add a subscript to $\mathcal{D}_{n,r}$ between $n$ and $r$ that indicates whether the data are generated from the null model $\Psi_0$ or alternative model $\Psi_1$. We also define criteria for the trial operating characteristics. Under $\Psi_1$ where $H_1$ is true, we want power in (\ref{eq:power.true}) for all experimental arms $j$ to be such that $\nu_j(\boldsymbol{\tau}(\mathcal{D}_{n})) \ge \Gamma_1$. We want the FWER in (\ref{eq:t1e.true}) to be such that $\nu_0(\boldsymbol{\tau}(\mathcal{D}_{n})) \le \Gamma_0$ under $\Psi_0$ where $H_0$ is true.

   We describe several additional inputs for Algorithm \ref{alg.star} below. The vector $\boldsymbol{\delta}_m$ denotes the non-inferiority margins that define the trial subhypotheses from Section \ref{sec:methods.1}. As demonstrated in Appendix A for the SSTARLET trial, the context surrounding the active sets $s \in \mathcal{S}$ and the constraint that $n_2 = c_2n$ ensure the sample sizes for all treatments in all trial stages are linear functions of $n$. This condition must be satisfied to apply Theorem \ref{thm1}, so it should be verified before adapting our SSD method to design other platform trials. 

\begin{algorithm}[H]
\caption{Sample size determination for the SSTARLET trial}
\label{alg.star}
\begin{algorithmic}[1]
\Require $n_a$, $\Psi_0$, $\Psi_1$, $R$, $\Gamma_0$, $\Gamma_1$, $p(\boldsymbol{\theta})$, $\boldsymbol{\delta}(\boldsymbol{\theta})$, $\boldsymbol{\delta}_m$, $\mathcal{S}$, $c_2$

\State Compute $\{\boldsymbol{\tau} (\mathcal{D}_{n_a,0,r})\}_{r=1}^R$ obtained with $\boldsymbol{\theta} \sim \Psi_0$.
\State Choose thresholds $\{\gamma_k\}_{k=1}^3$ and $\kappa_1$ such that $\widehat{\nu}_0\left(\{\boldsymbol{\tau} (\mathcal{D}_{n_a,0,r})\}_{r=1}^R\right) \le \Gamma_0$.
\State Compute $\{\boldsymbol{\tau} (\mathcal{D}_{n_a,1,r})\}_{r=1}^R$  obtained with $\boldsymbol{\theta} \sim \Psi_1$.
\State If $\widehat{\nu}_j\left(\left\{\boldsymbol{\tau}\left(\mathcal{D}_{n_a,1,r}\right)\right\}_{r=1}^R\right) > \Gamma_1$ for all $j \in \{1, 2, 3\}$, choose $n_b < n_a$. If not, choose $n_a > n_b$.
\State Compute $\{\boldsymbol{\tau} (\mathcal{D}_{n_b,1,r})\}_{r=1}^R$  obtained with $\boldsymbol{\theta} \sim \Psi_1$.
\For{$d \in \{1,\dots,R\}$}
    \For{$i \in \{1,\dots, L\}$}
        \State Let the sample $\mathcal{D}_{n_a,1,r}$ correspond to the $d$th order statistic of $\left\{\text{logit}(\tau_i (\mathcal{D}_{n_a,1,r}))\right\}_{r=1}^R$
        \State Pair the $d$th order statistics of $\left\{\text{logit}(\tau_i (\mathcal{D}_{n_a,1,r}))\right\}_{r=1}^R$ and $\left\{\text{logit}(\tau_i (\mathcal{D}_{n_b,1,r}))\right\}_{r=1}^R$ with linear \newline \hspace*{25pt} approximations to obtain $\text{logit}(\widehat{\tau}_i (\mathcal{D}_{n,1,r}))$ for new $n$ values.
    \EndFor
\EndFor
\State Obtain $\left\{\widehat{\boldsymbol{\tau}}(\mathcal{D}_{n,1,r})\right\}_{r=1}^R$ as the inverse logits of the estimates $\text{logit}(\widehat{\boldsymbol{\tau}} (\mathcal{D}_{n,1,r}))$.
\State Find $n$, the smallest $n \in \mathbb{Z}^+$ such that $\widehat{\nu}_j\left(\{\widehat{\boldsymbol{\tau}}(\mathcal{D}_{n,1,r})\}_{r=1}^R\right) \ge \Gamma_1$ for all $j \in \{1, 2, 3\}$.\\
\Return $n$ and $\{\gamma_1, \gamma_2, \gamma_3, \kappa_1\}$ as the recommended interim sample size and decision thresholds.
\end{algorithmic}
\end{algorithm}

We now elaborate on several steps of Algorithm \ref{alg.star}. In Line 2, we choose suitable values for the decision thresholds to ensure the FWER estimate in (\ref{eq:t1e.est}) is at most $\Gamma_0$. Because we estimate the sampling distribution of $\boldsymbol{\tau}(\mathcal{D}_n)$ via simulation, these recommended thresholds account for the dependence between endpoints and across stages of the trial along with the likelihood of observing each active set $s \in \mathcal{S}$. We note that all limiting slopes in Theorem \ref{thm1} are zero when $\boldsymbol{\theta} \sim \Psi_0$ is such that $\delta^*_k = \delta_{m,k}$ for all trial subhypotheses $k = 1, 2, 3$. The thresholds from Line 2 therefore approximately maintain the desired FWER across all sample sizes $n$ for such models $\Psi_0$.  

Line 5 of Algorithm \ref{alg.star} uses synthetic trial data to compute logits of posterior probabilities under $\Psi_1$: $\text{logit}(\tau_i(\mathcal{D}_{n, 1, r}))$ for $i =1, \dots, L$. In Lines 6 to 9, we separately construct linear approximations for each element in $\boldsymbol{\delta}(\boldsymbol{\theta})$ that are based on those logits. We use these linear approximations to estimate logits of posterior probabilities for new values of $n$ as $\text{logit}(\widehat{\tau}_i(\mathcal{D}_{n, 1, r}))$. We place a hat over the $\widehat{\tau}_i(\mathcal{D}_{n, 1, r})$ here to convey that this logit was estimated using a linear approximation instead of synthetic trial data. 

To maintain the proper level of dependence in the joint sampling distribution of $\boldsymbol{\tau}(\mathcal{D}_{n})$, we group the linear functions from Line 9 across $i \in \{1, \dots, L\}$ based on the sample $\mathcal{D}_{n_a, 1, r}$ that defined the linear approximations. Because Theorem \ref{thm1} ensures that the limiting derivatives of the logits of $\tau^{_{(n)}}_{i, u_i}$ and $\tau^{_{(n)}}_{i, u_i}~|~\{\tau^{_{(n)}}_{i', u_{i'}}\}_{i' = 1}^{i-1}$ from the proxy sampling distribution are the same, we use the marginal sampling distributions of posterior probabilities to estimate slopes for the conditional logits. Given the linear trend in the conditional proxy sampling distribution quantiles from Theorem \ref{thm1}, it is reasonable to construct these linear approximations based on order statistics from estimates of the true sampling distributions. 

In Line 11 of Algorithm \ref{alg.star}, we find the smallest value of $n$ such that the power estimates in (\ref{eq:power.est}) based on $\left\{\widehat{\boldsymbol{\tau}}\left(\mathcal{D}_{n,1,r}\right)\right\}_{r=1}^R$ from Line 10 are at least $\Gamma_1$ for all experimental arms $j$. Sample sizes for all arms in all stages of the trial can obtained using the relationships defined in Appendix A. Importantly, the interim posterior probabilities in $\widehat{\tau}_i(\mathcal{D}_{n, 1, r})$ depend on $n$, and therefore the corresponding active set $s^{(r)}$--i.e., the set of arms that remain under evaluation after the interim--also changes with $n$. To accommodate this, we explicitly enumerate all possible active sets $s^{(r)}\in\mathcal{S}$ for each simulation replicate $r$, allowing the active set to vary with both sample size and decision thresholds. Since our SSD method models the sampling distribution of $\boldsymbol{\tau}(\mathcal{D}_{n})$ independently of any specific interim decision path, it can easily accommodate changes in the active sets across sample sizes and thresholds.

Unlike most simulation-based approaches, which condition on a fixed sequence of interim decisions and must re-run simulations for every change in the sample size or decision threshold, our method can accommodate such changes without needing to run new simulations. Although Algorithm \ref{alg.star} requires evaluating a more complex sampling distribution--one that accounts for all possible active sets--it does so at only two values of $n$, offering a computationally efficient yet flexible approach. In Section \ref{sec:sim}, we demonstrate the performance and benefits of this streamlined SSD method for the SSTARLET trial design.



\begin{table}[ht]
\caption{Marginal probabilities for binary outcomes in various design scenarios.}
\label{tab:scenarios}
\centering
\renewcommand{\arraystretch}{1.2}
\footnotesize
\begin{tabular}{|l|c|PcPc|}
\hline
\textbf{Rate Outcome} & \textbf{Reference} & \multicolumn{4}{c|}{\textbf{Experimental Treatment}} \\
\cline{3-6}
& & \textbf{1: Clearly Acceptable} & \textbf{2: Acceptable} & \textbf{3: Barely Acceptable} & \textbf{4: Unacceptable} \\
\hline
AE              & 2\%  & 2\%  & 3\%  & 5\%  & 6\% \\
Completion      & 75\% & 75\% & 72\% & 70\% & 65\% \\
Non-tolerability & 25\% & 25\% & 28\% & 30\% & 35\% \\
\hline
\end{tabular}
\end{table}

\section{Simulation Study}\label{sec:sim}

We now apply our proposed approach to determine the optimal sample size and decision thresholds for the SSTARLET trial. We begin by considering four design scenarios for a given experimental treatment, summarized in Table~\ref{tab:scenarios}. The first three scenarios represent configurations under which the treatment is non-inferior to the control (i.e., the alternative hypothesis \(H_1\) holds), while the fourth scenario corresponds to inferiority (i.e., the null hypothesis \(H_0\) holds). Our objective is to select the smallest final sample size such that the power to correctly declare non-inferiority at the final analysis under Scenario 1 (Clearly Acceptable) is at least \(\Gamma_1 = 0.95\) for each treatment. The interim and final decision thresholds, $\boldsymbol{\gamma} = (0.2, 0.5, 0.5)$ and $\boldsymbol{\kappa} = (0.975, 0.99, 0.99)$, were selected as follows: $\kappa_1$ was chosen to control the FWER under Scenario 4 (Unacceptable) at $\Gamma_0 = 0.05$, while the $\boldsymbol{\gamma}$ values and the thresholds $\kappa_2$ and $\kappa_3$ were selected heuristically to ensure desirable interim stopping behavior and to balance the risk of incorrect versus correct final decisions. Probabilities of concluding non-inferiority under Scenario 2 (Acceptable) and Scenario 3 (Barely Acceptable) are examined in sensitivity analyses to assess the behavior of the design across a range of plausible outcomes.

\begin{figure}[H]
    \centering
    \includegraphics[width=0.75\textwidth]{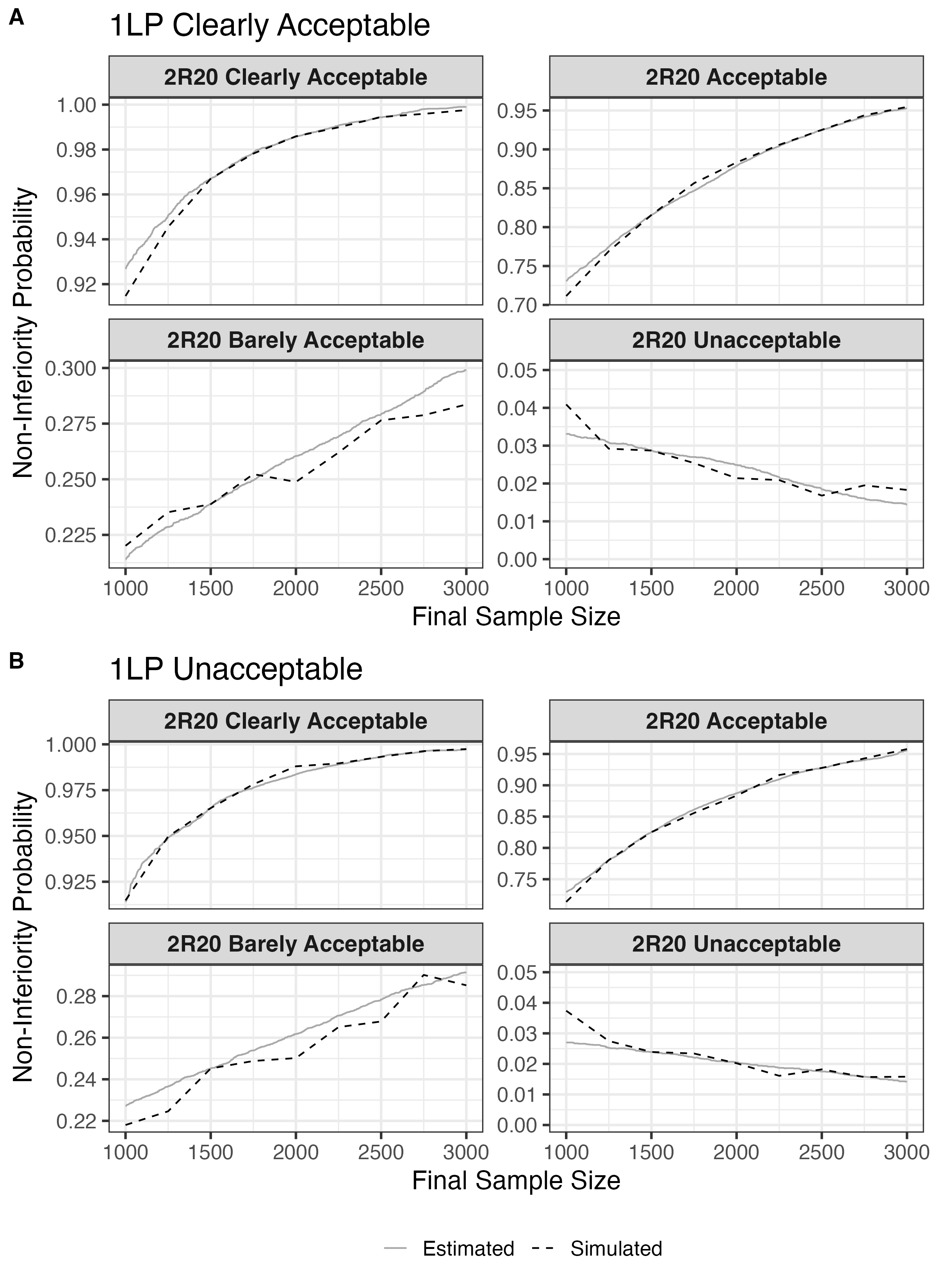}
    \caption{Estimated (gray solid) vs simulated (black dotted) non-inferiority probabilities for 2R20 across four design scenarios, under 1LP Clearly Acceptable (A) and 1LP Unacceptable (B).}
    \label{fig:2r20vary}
\end{figure}

Because the interim decisions for each treatment arm influence how remaining participants are allocated, the performance of any one treatment cannot be assessed in isolation. In particular, whether another experimental treatment is dropped or retained at the interim analysis affects the sample size available for the remaining arms. Therefore, to accurately evaluate the overall operating characteristics of the trial, we must consider combinations of outcome scenarios across all treatment arms. Specifically, we assess each of the four possible profiles (Clearly Acceptable, Acceptable, Barely Acceptable, and Unacceptable) for the third unspecified experimental treatment, referred to as Trt3, under all combinations of 2R20 and 1LP being either Clearly Acceptable or Unacceptable. This yields a total of 16 composite scenarios to evaluate non-inferiority probability for Trt3. 

For the remaining arms, we examine non-inferiority conclusions for 2R20 across the four design scenarios under each of the two possible configurations of 1LP (Clearly Acceptable or Unacceptable), and analogously for 1LP across configurations of 2R20. This results in 8 evaluation scenarios for each of 2R20 and 1LP. We do not condition on the profile of Trt3 when analyzing 2R20 or 1LP, as the addition of Trt3 occurs at the interim analysis trigger and does not influence allocation or the probability of early stopping for these arms. Although we must consider all the aforementioned scenarios to assess the SSTARLET design, we conduct SSD via Algorithm \ref{alg.star} with a model $\Psi_1$ such that all experimental arms are Clearly Acceptable and a model $\Psi_0$ such that all arms are Unacceptable. We elaborate on this choice for $\Psi_1$ shortly.

\begin{figure}[H]
    \centering
    \includegraphics[width=0.75\textwidth]{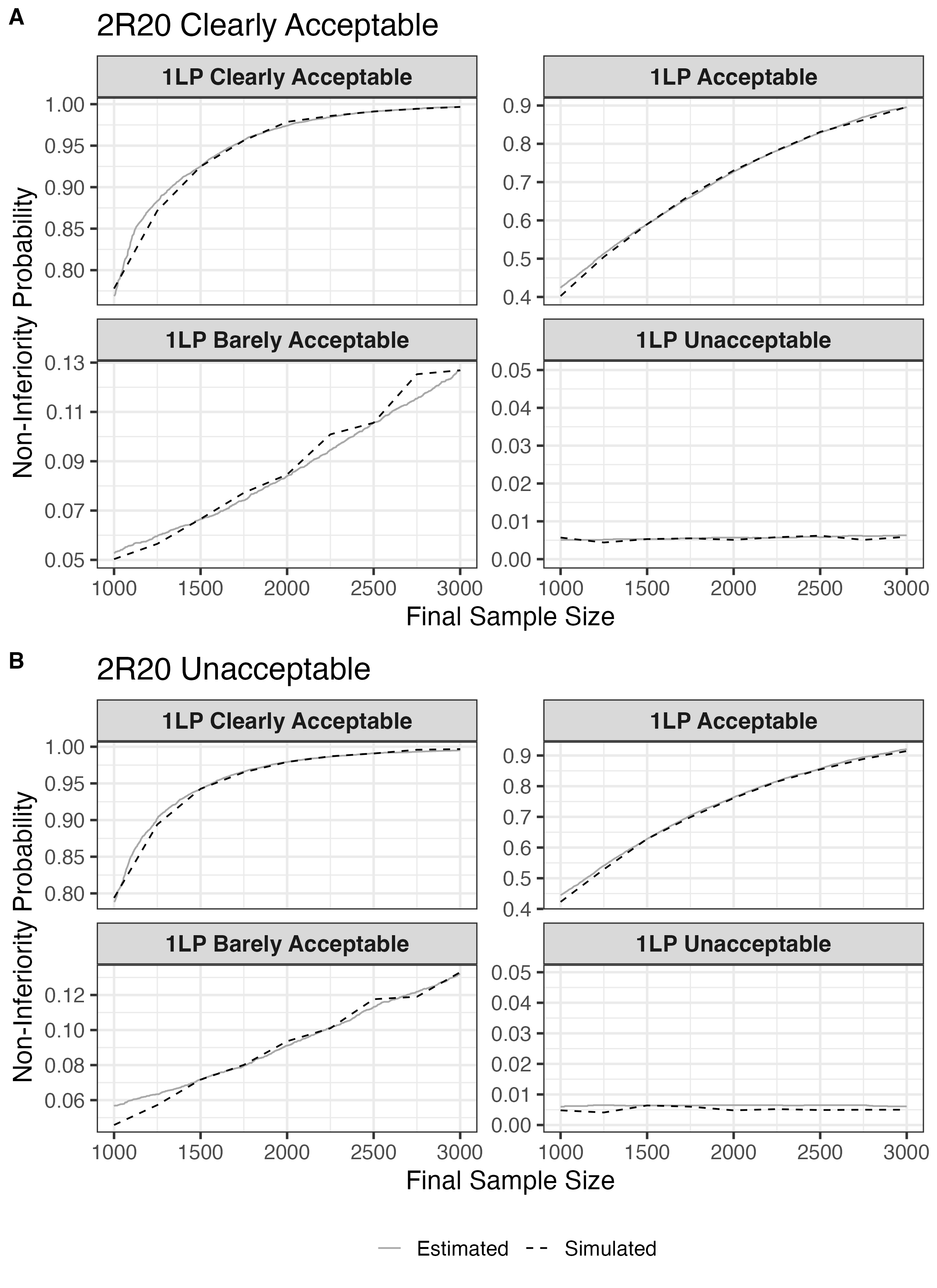}
    \caption{Estimated (gray solid) vs simulated (black dotted) non-inferiority probabilities for 1LP across four design scenarios, under 2R20 Clearly Acceptable (A) and 2R20 Unacceptable (B).}
    \label{fig:1hpvary}
\end{figure}

\begin{figure}[H]
    \centering
    \includegraphics[width=\textwidth]{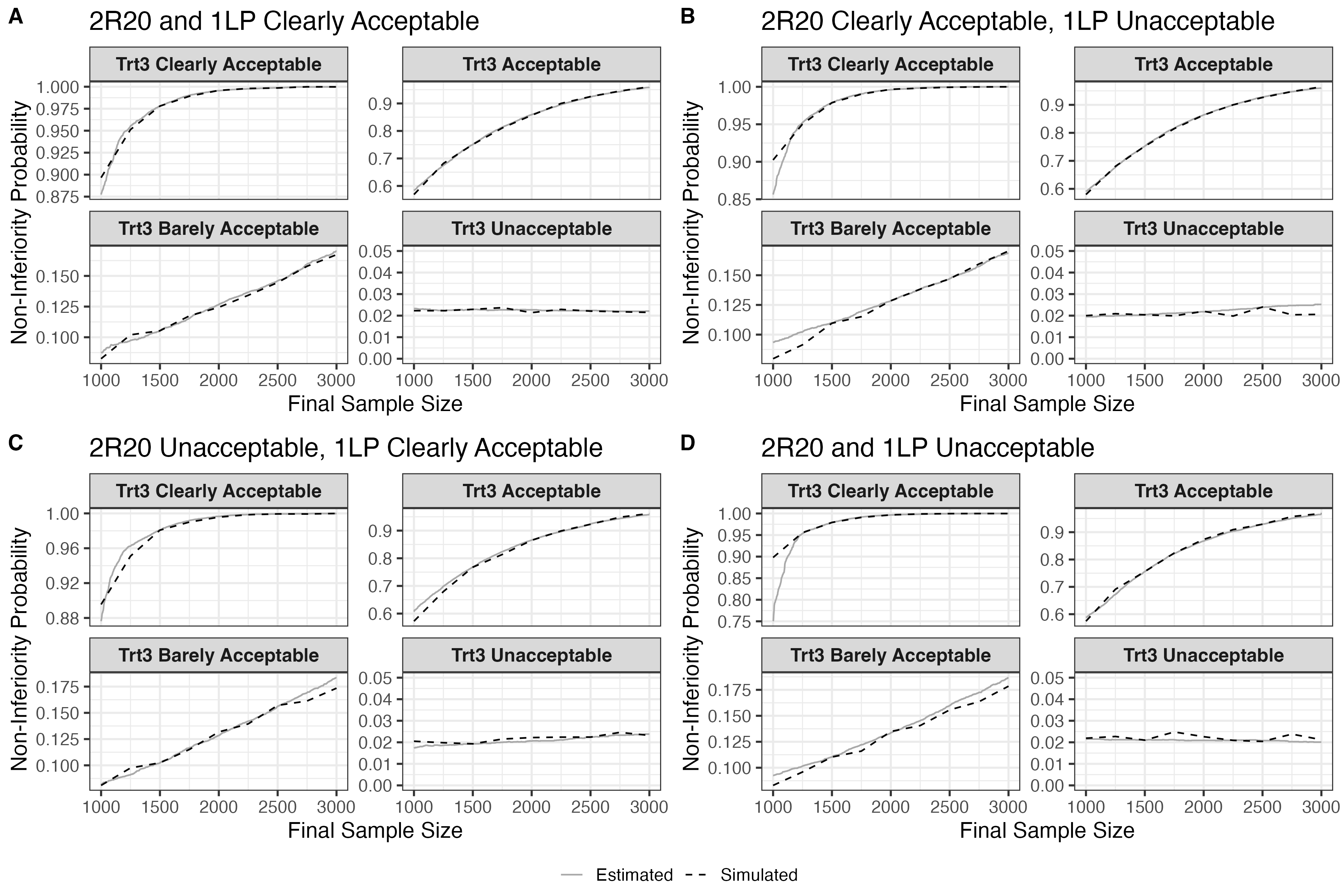}
    \caption{Estimated (gray solid) and simulated (black dotted) non-inferiority probabilities for Trt3 under four setting combinations: (A) both 2R20 and 1LP are Clearly Acceptable, (B) 2R20 is Clearly Acceptable while 1LP is Unacceptable, (C) 2R20 is Unacceptable while 1LP is Clearly Acceptable, (D) both 2R20 and 1LP are Unacceptable.}
    \label{fig:trt3subset}
\end{figure}

Figures~\ref{fig:2r20vary}, \ref{fig:1hpvary}, and \ref{fig:trt3subset} visualize the probability of concluding non-inferiority at the final analysis across the four design scenarios defined in Table~\ref{tab:scenarios}, plotted against the final sample size. This final sample size is $n_2 = 2.5n_1$ (i.e., $c_2 = 2.5$). For Scenarios 1, 2, and 3, this probability corresponds to power for arm $j$, while for Scenario 4 it represents the treatment-specific type I error rate. The gray solid curves show estimates obtained using our SSD procedure, which linearly approximates the logits of posterior probabilities based on samples from two anchor values of the interim sample size, \( n_a = 600 \) and \( n_b = 1000 \), following Lines 3–10 of Algorithm~\ref{alg.star}. In contrast, the black dotted curves reflect conventional simulation-based estimates, computed by independently generating synthetic datasets and evaluating posterior probabilities at a grid of values for the interim sample size \( n_1 \in \{400, 500, \dots, 1200\} \) corresponding to a final sample size of \( n_2 \in \{1000, 1250, \dots, 3000\} \) for each scenario. We used $R = 10,000$ replicates for all sampling distribution estimates in both methods.

While the simulation-based curves exhibit some simulation variability, they serve as approximate surrogates for the true design operating characteristics. Across all four scenarios, we generally observe strong agreement between the gray solid (SSD-based) and black dotted (simulation-based) curves, suggesting that the SSD procedure closely tracks the empirical sampling distribution of posterior probabilities. In subplots B and D of Figure \ref{fig:trt3subset}, there are some discrepancies between the estimated and simulated curves for smaller sample sizes when Trt3 is Clearly Acceptable. As illustrated in Appendix C of the online supplement, the agreement between the estimated and simulated curves in these settings substantially improves as more samples are drawn from the posterior distribution. 

Notably, power is slightly improved in scenarios where one or more experimental arms are dropped at the interim analysis. This arises because the remaining arms are allocated a larger proportion of the second-stage sample, enhancing precision and increasing the probability of meeting the non-inferiority criterion. We also note that power for 2R20 is substantially improved compared with that of 1LP due to the incorporation of a MAP prior, which contributes additional information to the posterior distribution. For Trt3, the modest increase in power reflects the allocation rule that directs 50\% of the second-stage sample to Trt3, regardless of which arms are retained. It took roughly 22 minutes on a high-computing server to estimate each gray solid curve in Figures \ref{fig:2r20vary}, \ref{fig:1hpvary}, and \ref{fig:trt3subset} when
approximating each posterior using 1000 draws. We considered
9 values of $n$ to simulate each black dotted curve in the figures, taking approximately 1 hour and 40 minutes using the same computing resources.

\begin{table}[ht]
\caption{Estimated and simulated non-inferiority probabilities at the final analysis across four outcome scenarios (columns) for each treatment and setting (grouped rows).}
\centering
\footnotesize
\begin{tabular}{LlPcPc}
\hline
\textbf{Setting} & \textbf{Source} & \textbf{Clearly Acceptable} & \textbf{Acceptable} & \textbf{Barely Acceptable} & \textbf{Unacceptable} \\
\hline
\multicolumn{6}{c}{\textbf{Non-Inferiority Probability for 2R20}} \\
\hline
1LP Clearly Acceptable & Estimated & 0.9761 & 0.8397 & 0.2473 & 0.0274 \\ 
& Simulated & 0.9752 & 0.8480 & 0.2338 & 0.0260 \\ 
1LP Unacceptable & Estimated & 0.9744 & 0.8526 & 0.2520 & 0.0228 \\ 
& Simulated & 0.9749 & 0.8589 & 0.2484 & 0.0233 \\ 
\hline
\multicolumn{6}{c}{\textbf{Non-Inferiority Probability for 1LP}} \\
\hline
2R20 Clearly Acceptable & Estimated & 0.9497 & 0.6442 & 0.0723 & 0.0054 \\ 
& Simulated & 0.9498 & 0.6560 & 0.0779 & 0.0050 \\ 
2R20 Unacceptable & Estimated & 0.9605 & 0.6847 & 0.0776 & 0.0058 \\ 
& Simulated & 0.9569 & 0.6855 & 0.0790 & 0.0058 \\ 
\hline
\multicolumn{6}{c}{\textbf{Non-Inferiority Probability for Trt3}} \\
\hline
\multirow{2}{=}{2R20 and 1LP\\Clearly Acceptable}  & Estimated & 0.9877 & 0.7997 & 0.1127 & 0.0226 \\ 
& Simulated & 0.9898 & 0.8004 & 0.1146 & 0.0199 \\ 
\multirow{2}{=}{2R20 Clearly Acceptable, 1LP Unacceptable} & Estimated & 0.9885 & 0.8033 & 0.1165 & 0.0210 \\ 
& Simulated & 0.9887 & 0.8005 & 0.1137 & 0.0224 \\ 
\multirow{2}{=}{2R20 Unacceptable, 1LP Clearly Acceptable} & Estimated & 0.9894 & 0.8108 & 0.1130 & 0.0197 \\ 
& Simulated & 0.9873 & 0.8044 & 0.1172 & 0.0236 \\ 
\multirow{2}{=}{2R20 and 1LP Unacceptable} & Estimated & 0.9891 & 0.8067 & 0.1193 & 0.0211 \\ 
 & Simulated & 0.9871 & 0.8059 & 0.1158 & 0.0189 \\ 
\hline
\end{tabular}

\label{tab:ni-prob-summary}
\end{table}

Our method recommended a final sample size of \( n_2 = 1685 \), with an interim analysis planned at \( n_1 = 674 \). To further assess the performance of the proposed approach, Table~\ref{tab:ni-prob-summary} presents the estimated non-inferiority probability at the final analysis across a range of design scenarios. These probabilities were first obtained using the linear approximation strategy from Algorithm~\ref{alg.star}, anchored at interim sample sizes \( n_a = 600 \) and \( n_b = 1000 \). Each scenario is shown with a corresponding pair of rows: the first reflects the SSD-based estimates, while the second provides results from direct simulation of the joint sampling distribution of \( \boldsymbol{\tau}(\mathcal{D}_n) \) at the recommended sample size. The strong agreement observed between the two rows within each pair demonstrates the accuracy and robustness of our approximation method. Moreover, the results confirm that the design achieves the desired power of \( \Gamma_1 = 0.95 \) for each treatment in Scenario 1. Power to conclude non-inferiority was tuned to the most conservative subscenario, where all three experimental treatments are Clearly Acceptable. In this setting that informed our choice for the model $\Psi_1$, no arms are likely to be dropped at the interim analysis, so the total sample size remains distributed across all arms, resulting in smaller sample sizes per arm. Among the treatments, power is lowest for 1LP: it receives slightly less allocation than the third experimental arm due to the timing of the interim analysis and allocation ratios, and--unlike 2R20--does not benefit from an informative prior. As a result, the sample size was selected to ensure that 1LP achieves at least 0.95 power under this edge-case configuration, thereby guaranteeing that the desired power is met in all other subscenarios as well.


\section{Discussion}\label{sec:disc}

In this paper, we proposed an efficient framework to assess the operating characteristics for Bayesian platform trials. This framework determines the minimum sample size required to satisfy power criteria for the experimental arms while bounding the FWER. The computational efficiency of our design framework is motivated by considering a proxy for the joint sampling distribution of posterior probabilities across multiple endpoints, trial stages, and interim decisions. We use our theoretical results pertaining to this large-sample proxy distribution to justify estimating true sampling distributions at only two sample sizes. Thus, our method drastically reduces the number of simulation replicates required to design Bayesian platform trials. 

This work was motivated by the design complexities of the SSTARLET trial, which include mid-trial addition of experimental arms, fixed delays between interim recruitment and analyses, variable randomization ratios induced by potential early stopping, and the use of informative priors. Nevertheless, our proposed methodology can be broadly applied to design Bayesian clinical trials with multiple experimental interventions.

Several directions for future research are available. First, in our application, we analyzed each binary outcome using a separate model for simplicity. A natural extension would be to jointly model dependent binary outcomes using Bayesian hierarchical models that account for outcome correlations without substantially increasing computational complexity. While the threshold $\kappa_1$ was explicitly tuned to control the FWER, other decision thresholds in our current approach were selected heuristically to achieve desirable properties. Future work will explore more principled procedures for selecting all decision thresholds, for instance using utility functions or constrained optimization, especially in settings where only a subset of null hypotheses is false.

This paper focused on a specific form of adaptive randomization in which, after an arm is dropped at interim, its allocation is re-distributed among the remaining arms. We also addressed the presence of a fixed delay between reaching the interim enrollment threshold and the availability of outcome data—a common challenge in trials with longer follow-up. Future research will extend this work to more general adaptive randomization strategies, including outcome-adaptive or covariate-adjusted designs, and evaluate how different delay structures influence trial operating characteristics. On the modeling side, incorporating MAP priors for clustered binomial outcomes would enable information borrowing across arms or historical studies while accounting for within-cluster correlation. This would involve extending the MAP framework—typically based on independent data—to a hierarchical model that captures intra-cluster variability, for instance via beta-binomial or logistic mixed-effects formulations. The resulting prior would reflect both between- and within-cluster uncertainty, improving inference in cluster randomized trial settings. Separately, computational enhancements such as implementing power priors via Sequential Monte Carlo methods could improve scalability in high-dimensional or adaptively updated Bayesian designs.

\section*{Supplementary Material}
These materials include additional context for the SSTARLET trial and a proof of Theorem \ref{thm1}. The code to implement our method is available online: \url{https://github.com/lmhagar/PlatformSSD}.

\section*{Acknowledgements}
LH acknowledges the support of a postdoctoral fellowship from the Natural Sciences and Engineering
Research Council of Canada (NSERC).  LM is supported by a Canadian Network in Statistics and Trials (CANSTAT) trainee award funded by Canadian Institutes of Health Research (CIHR) grant \#262556. Shirin Golchi acknowledges support from NSERC, Canadian Institute for Statistical Sciences (CANSSI), Fonds de recherche du Qu\'ebec - Sant\'e (FRQS) and Fonds de recherche du Qu\'ebec - Nature et technologies (FRQNT). DM is a Canada Research Chair (Tier 1). SSTARLET is supported by the Clinical Trials Fund of CIHR. DM also holds affiliations with the Montreal Chest Institute at the McGill University Health Center, the Research Institute of the McGill University Health Center, and the Department of Medicine of the Faculty of Medicine and Health Sciences at McGill University.

\newpage 
\bibliographystyle{apalike}
\bibliography{bib}

\end{document}